\documentclass[aps,preprint,pra,floatfix,showpacs]{revtex4}
\usepackage{graphicx}

\begin{document}
\title{Substrate enhanced superconductivity in Li-decorated graphene}
\author{T. P. Kaloni$^1$, A. V. Balatsky$^{2,3,4}$, and U.\ Schwingenschl\"ogl$^{1}$}
\email{udo.schwingenschlogl@kaust.edu.sa}
\affiliation{$^1$PSE Division, KAUST, Thuwal 23955-6900, Kingdom of Saudi Arabia\\
$^2$Theoretical Division, Los Alamos National Laboratory, Los Alamos, New Mexico 87545, USA\\
$^3$Center for Nanotechnologies, Los Alamos National Laboratory, Los Alamos, New Mexico 87545, USA\\
$^4$Nordita, KTH Royal Institute of Technology and Stockholm University, Roslagstullsbacken 23, SE-106 91 Stockholm Sweden}

\begin{abstract}
We investigate the role of the substrate for the strength of the electon phonon coupling
in Li-decorated graphene. We find that the interaction with a $h$-BN substrate leads to
a significant enhancement from $\lambda_0=0.62$ to $\lambda_1=0.67$, which corresponds to
a 25\% increase of the transition temperature from $T_{c0}=10.33$ K to $T_{c1}=12.98$ K.
The superconducting gaps amount to 1.56 meV (suspended) and 1.98 meV (supported). These 
findings open up a new route to enhanced superconducting transition temperatures in
graphene-based materials by substrate engineering.
\end{abstract}

\pacs{74.78.Db, 63.20.Dj, 81.05.Uw, 31.15.Ar}
\maketitle

\section{Introduction}
Recent observations in alkali-doped graphene have opened exciting venues to superconductivity
accomplished by doping \cite{mouri}. Most theoretical estimates of the electron-phonon
coupling so far have assumed suspended graphene as a base, since this geometry makes
calculations more direct and less computationally costly. However, most of the engineered
superconducting graphene samples use a substrate. Hence, it is important to characterize
the role of the substrate on the superconductivity in atomically thin graphene. We have
performed a first-principles study of the role of the substrate on the phonon spectrum
and the electron-phonon coupling and find not only that the interaction with the substrate is
relevant but that in the case of a $h$-BN substrate the electron-phonon coupling can be
enhanced by as much as 9\% so that $T_c$ can be expected to reach 12.98 K, a 25\% increase.
This observation points to a new direction in the search for novel superconducting materials:
substrate-engineered superconductivity, where the nascent superconducting states are
significantly enhanced by the coupling to a properly chosen substrate.

Graphite intercalated compounds are characterized by a nearly free electron band,
which upon increased doping crosses the Fermi energy ($E_F$). Empirically, there are
intercalated compounds that exhibit superconductivity with a transition temperature
of a few to about 10 K. An empirical correlation between the crossing of the chemical
potential and the onset of superconductivity was first put forward in Ref.\ \cite{Littlewood}
and subsequently was called the ``Cambridge criterion" \cite{balatsky}. Superconductivity
in Ca-intercalated bilayer graphene has been predicted with a sizable $T_c = 11.5$ K by
analyzing this criterion in Refs.\ \cite{balatsky,jishi}. Recently, the prediction
has been verified experimentally for Ca-intercalated graphene on either the Si or the
C face of a SiC substrate, finding $T_c=7$ K \cite{Li}. Experimentally, it also has
been observed that KC$_8$ graphite \cite{valla} and K-intercalated few layer graphene
on SiC are superconducting \cite{jacs}, where theoretical arguments for superconductivity
in the latter material have been presented in Ref.\ \cite{kaloni-epl}. To complete the
list of superconducting C allotropes we also mention that undoped single and multiwall nanotubes
exhibit superconductivity with a sizable $T_c\sim 10$ to 12 K \cite{Tang,Takesue}. Today,
the highest value of $T_c=38$ K is observed experimentally in Cs$_3$C$_{60}$ \cite{Alexey}.

It was already pointed out that not all intercalants lead to an enhanced $T_c$ in graphite
\cite{Littlewood}. A high $T_c$ can be obtained when the distance between the intercalated
atom and the graphene plane is small so that the deformation potential is large \cite{Boeri}.
Our observations are consistent with this mechanism: The distance between the intercalant
and the graphene plane is 2.62 \AA, 2.47 \AA, and 2.26 \AA\ for non-superconducting BaC$_6$
\cite{Boeri1}, superconducting SrC$_6$ with $T_c=1.65$ K \cite{Boeri1}, and superconducting
CaC$_6$ with $T_c=11.5$ K \cite{Emery,Ellerby}, respectively, see Table I. 

\begin{table*}[t]
\begin{tabular}{|c|c|c|}
\hline 
compound & distance &$T_c$ \tabularnewline  
\hline
BaC$_6$  & 2.62 \AA      &0 K\tabularnewline
\hline
SrC$_6$  &2.47  \AA     &1.65 K\tabularnewline
\hline
CaC$_6$  &2.26   \AA    &11.5 K  \tabularnewline  
\hline
\end{tabular}
\caption{Compound, perpendicular distance of the intercalated atom from the center of the
C hexagon, and superconducting transition temperature.}
\end{table*}

In this paper we report a first-principles study of the electron-phonon coupling to estimated
the values of $\lambda$ and $T_c$ for Li-decorated suspended graphene and Li-decorated
graphene on a $h$-BN substrate. We show that the presence of the substrate enhances the
electron-phonon coupling and superconducting transition temperature, which reflects a
significant impact of the interaction of the electronic states with the substrate on the
phonon mediated superconductivity in doped graphene.

\section{Results and discussion}
The unit cell of Li-decorated monolayer graphene comprises 6 C atoms and 1 Li atom in a
$\sqrt3\times\sqrt3R30{^\circ}$ geometry, where the Li atom lies above the
center of the C hexagon in a distance of 1.76 \AA, slightly smaller than the value reported
in Ref.\ \cite{mouri}. The possible reason for the latter is inclusion of the van der Waals
interaction in our calculations, which is expected to provide a correct interlayer spacing.
The structural arrangements of Li-decorated graphene suspended and supported by a $h$-BN
substrate are presented in Figs.\ 1(a) and 1(b). The electronic band structures obtained
for $\sqrt3\times\sqrt3R30{^\circ}$ suspended graphene without and with Li-decoration
are shown in Figs.\ 2(a) and 2(b). It is well known that the C $\pi$ and $\pi^*$ orbitals
form a Dirac cone at the Fermi energy. Due to Brillouin zone backfolding,
the Dirac cone appears at the $\Gamma$-point and not at the K-point as in the case of the
primitive unit cell of graphene.

\begin{figure*}[t]
\includegraphics[width=0.4\columnwidth]{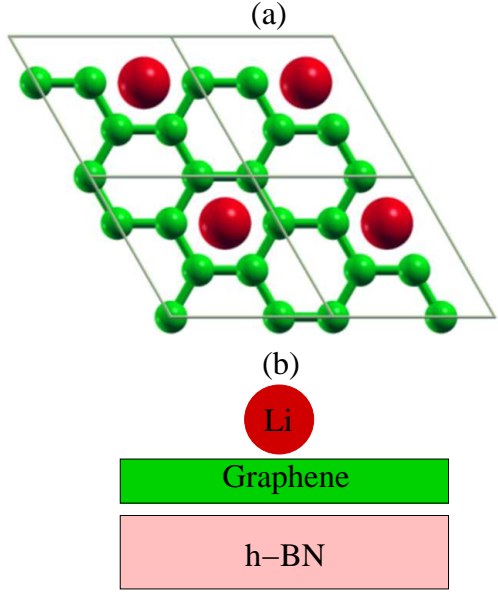}
\caption{Crystal structure of Li-decorated graphene (a) suspended and (b) supported by
a $h$-BN substrate.}
\end{figure*}

The electronic band structure of Li-decorated graphene is found to be modified significantly
as compared to that of pristine graphene. The nearly free electron Li $s$ band crosses the
Fermi level, due to charge transfer from Li to C.
As a consequence, the ``Cambridge criterion" is satisfied and the system should
be a superconductor. We will comment later on this phonomenon by analyzing the strength of
the electron-phonon coupling. A gap of 0.38 eV opens 1.56 eV below the Fermi level, as to be
expected \cite{kaloni-cpl,Farjam}. In Fig.\ 2(b) the partially occupied parabolic bands
indicated by arrows are due to Li $s$ states, compare
Figs.\ 2(a) and 2(b). It has been reported that the carrier density in Li-decorated
monolayer and Li-intercalated multilayer graphene with and without substrate can differ by
a factor of 100 from that of pristine graphene \cite{kaloni-cpl}.

\begin{figure*}[t]
\includegraphics[width=0.99\columnwidth]{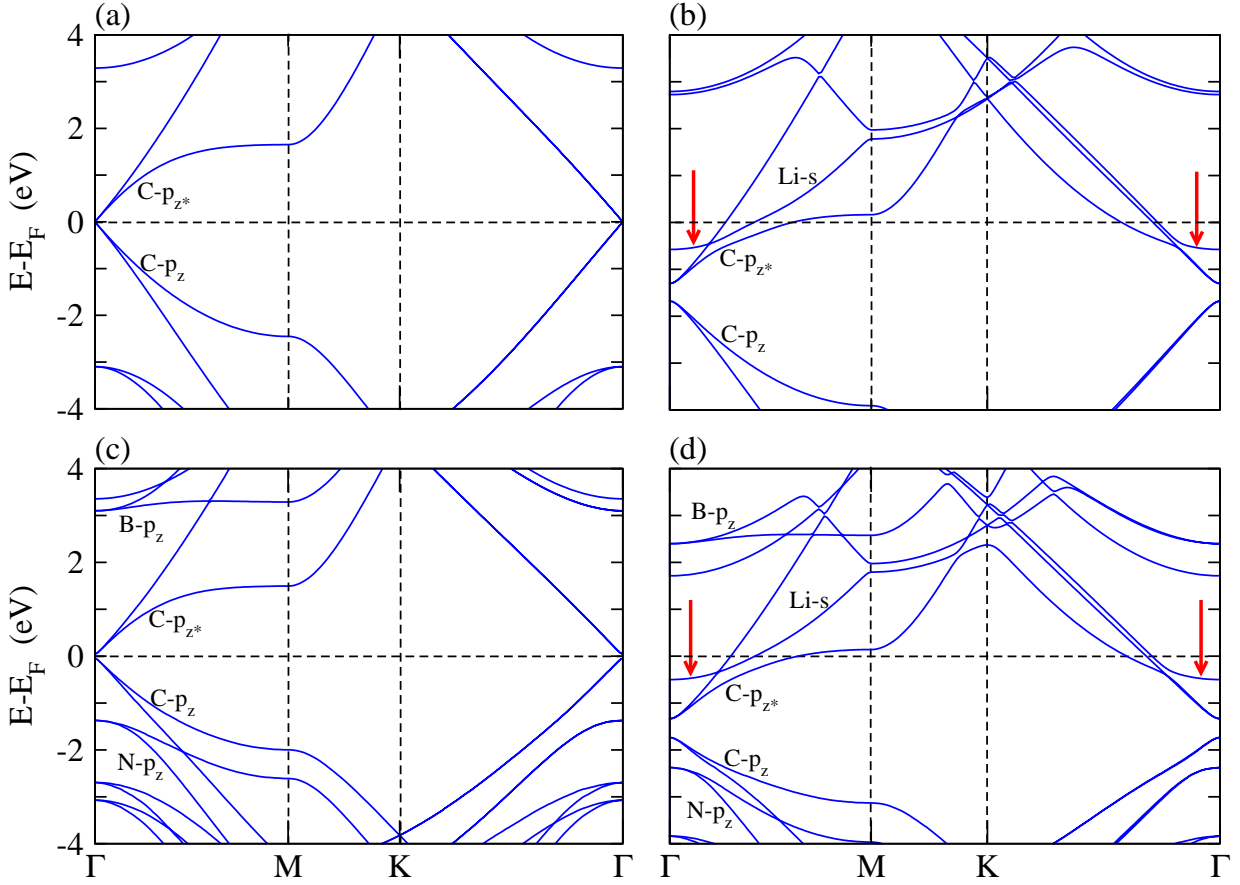}
\caption{Electronic band structure of (a) suspended C$_6$, (b) suspended C$_6$Li, (c)
C$_6$ on $h$-BN, and (d) C$_6$Li on $h$-BN. Li $s$ bands crossing the Fermi level are indicated
by arrows.}
\end{figure*}

For Li-decorated graphene on $h$-BN, see Fig.\ 1(b), the separation between graphene and
the substrate is found to be 3.39 \AA, which is close to the values for superlattices of
graphene and $h$-BN as well as graphene on a $h$-BN substrate. The perpendicular distance
of the Li atom to the graphene plane is 1.77 \AA. The lost sublattice symmetry (only each
third C hexagon is occupied by a Li atom) is responsible for a band gap of 90 meV. This value
agrees well with previous reports \cite{Quhe,Naveh,kaloni-jmc}, which also applies to the
fact that the B and N states appear far away from the Fermi level. The electronic band
structure in Fig.\ 2(d) clearly shows that a nearly free electron Li $s$ band crosses the
Fermi level, satisfying the ``Cambridge criterion" and thus pointing to superconductivity
in the system. The nature and magnitude of the gap at the $\Gamma$-point just below the
Fermi level are similar to Fig.\ 2(b).

At this point we assume that the basic mechanism of the superconductivity in the
suspended and supported cases is the same as in Ca-intercalated graphene, i. e.,
electron-phonon driven pairing \cite{balatsky}. It has been proposed that the dopant-induced
soft phonon modes contribute substantially to the electron-phonon coupling
\cite{Mazin,Littlewood} and it is known that the motion of the adatom is responsible for
about half of the coupling, while the other half is due to the C atoms
\cite{Calandra,Sanna,Rosenmann}. The presence of the Li $s$ states around the Fermi energy
alone cannot be sufficient to give a large electron-phonon coupling \cite{mouri}, but
the coupling to the out-of-plane C vibrations plays an important role due to
transitions between the C $\pi^*$ and Li $s$ states. The Li $s$ band enhances the coupling
\cite{Boeri} and, hence, the transition temperature. For this reason, we calculate
the phonon dispersion, see Fig.\ 3(a), and $\alpha^2F(\omega)$, see Fig.\ 3(b), and estimate
the strength of the electron-phonon coupling $\lambda$ using Eq.\ (2).

For the phonon dispersion of Li-decorated suspended graphene we find that most modes between
300 cm$^{-1}$ and 500 cm$^{-1}$ are due to a mixture of Li and out-of-plane C vibrations.
The pure out-of-plane modes appear from 500 cm$^{-1}$ to 900 cm$^{-1}$ and higher energy
C-C stretching modes from 900 cm$^{-1}$ to 1515 cm$^{-1}$. The modes from 300 cm$^{-1}$ to
500 cm$^{-1}$ are responsible for the electron-phonon coupling. This also can be seen from
$\alpha^2F(\omega)$ as addressed in Fig.\ 3(b). Experimentally, for pristine graphene the
frequency of the G-mode is 1580 cm$^{-1}$ \cite{mouri1}, which softens to 1515 cm$^{-1}$
under Li decoration. The softening can be attributed to charge transfer from Li to graphene
and the induced stronger electron-phonon coupling, in agreement with findings for the
molecular/atomic charge transfer in graphene \cite{Rao,carbon}. We obtain for the
electron-phonon coupling $\lambda=0.62$ and estimate for the superconducting transition
temperature $T_c=10.33$ K. This value is slightly higher than that of Ref.\ \cite{mouri},
since we take into account the van der Waals interaction to achieve an accurate distance
to the Li atom.

\begin{figure*}[t]
\includegraphics[width=0.99\columnwidth]{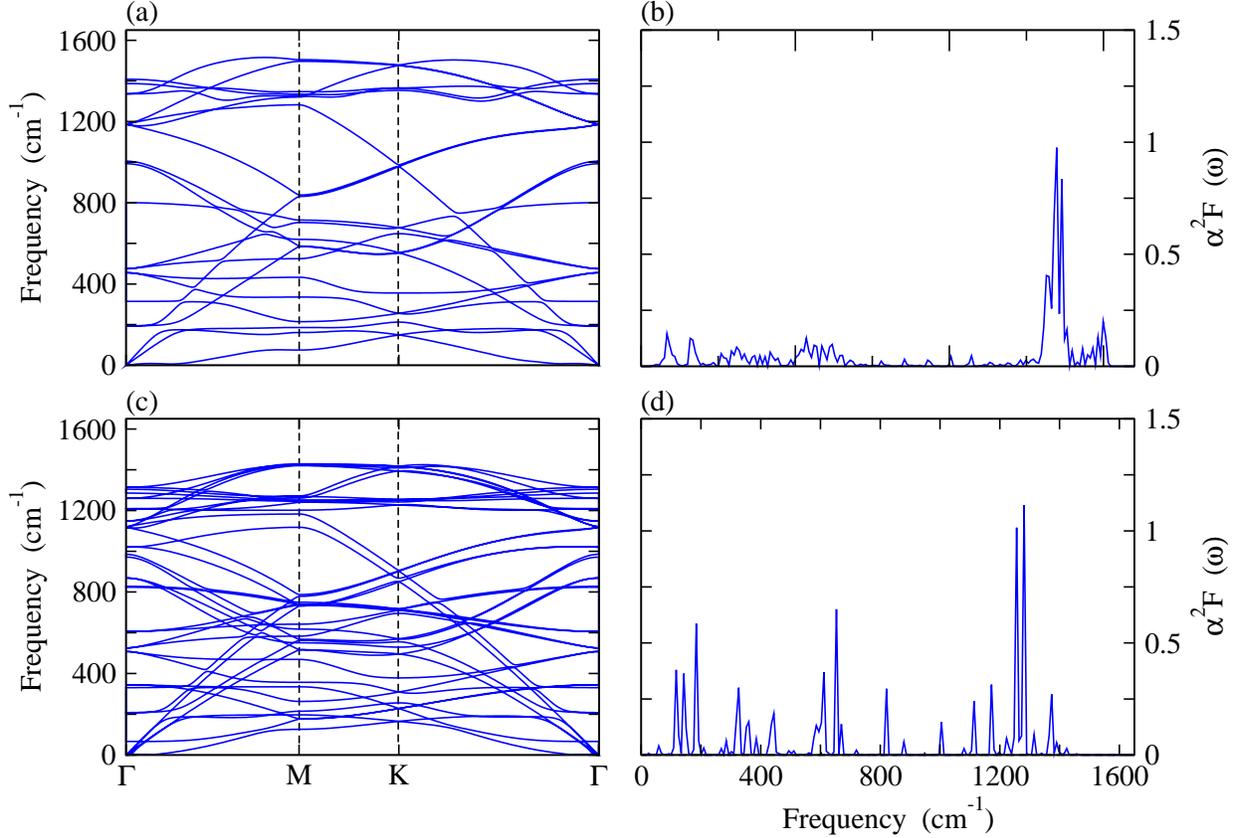}
\caption{\label{fig2} Electron-phonon dispersion of Li-decorated graphene (a) suspended and
(c) supported by a $h$-BN substrate. (b,c) Corresponding Eliashberg functions.}
\end{figure*}

Our central result is that an enhancement of the superconductivity in Li-decorated graphene can
be achieved by the application of a $h$-BN substrate. The phonon modes between 100 cm$^{-1}$
and 300 cm$^{-1}$ at the $\Gamma$-point are attributed to the Li vibrations, out-of-plane C
vibrations, and $h$-BN substrate, see Fig.\ 3(c). There are also substrate modes around
850 cm$^{-1}$ as well as higher energy modes between 900 cm$^{-1}$ and 1430 cm$^{-1}$, which
are due to both the substrate and C-C stretching. The softening of the modes as compared to
the suspended system is due to the interaction with the substrate, as observed experimentally,
for example, in graphite supported by Ni(111) \cite{new}. The modes in the range from
100 cm$^{-1}$ to 300 cm$^{-1}$ are responsible for a shift in $\alpha^2F(\omega)$ see
Fig.\ 3(d), and enhancement of $\lambda$ and $T_c$. We obtain $\lambda=0.67$ (as compared
to $\lambda=0.62$ in the suspended system) and thus a higher $T_c=12.98$ K, which is a
25\% increase with respect to the suspended system. The clearly indicates that one can take
full advantage of the substrate to boost $T_c$. Similarly, it has been observed experimentally
in the FeSe$_{0.5}$Te$_{0.5}$ superconductor that $T_c$ is enhanced by 15\% if the material
is supported \cite{Johnson}, while the details of the band structure are very different
in the present material. The most likely reason for the obtained enhancement of $T_c$ by
the application of a $h$-BN substrate are stronger spin fluctuations due to the lattice
mismatch of 1.4\%. Finally, we estimate the superconducting gap $\Delta_{sc}$ by the
relation $1.75 k_BT_c = \Delta_{sc}$ \cite{Parker}, where $k_B$ is the Boltzmann constant.
We obtain for Li-decorated suspended and supported graphene, respectively, values of 1.56 meV
and 1.98 meV.

\section{conclusion}
In conclusion, using density functional theory we have investigated the role of the substrate
for the electron-phonon coupling in Li-decorated suspended and supported graphene. We find
that the interaction with a $h$-BN substrate significantly enhances the electron-phonon coupling
to $\lambda=0.67$ as compared to $\lambda=0.62$ in the suspended case. The transition
temperature thus is enhanced by 25\% to 12.98 K. The superconducting gap for the suspended
and supported systems is found to be 1.56 meV and 1.98 meV, respectively. Our results show
that graphene-based nanomaterials can be tailored by properly choosing the substrate to
robustly increase the superconducting transition temperature.

\begin{acknowledgments}
We thank G.\ Profeta for fruitful discussions. This work is supported by US DOE, ERC-DM-321031,
and VR.
\end{acknowledgments}

\section{Appendix}
\subsection{Computational details}
All the results are obtained from density functional theory in the local density approximation.
The van der Waals interaction is taken into account via Grimme's scheme \cite{grime}. We use
the Quantum-ESPRESSO code \cite{paolo} with norm-conserving pseudopotentials and a plane wave
cutoff energy of 70 Ryd. A Monkhorst-Pack $32\times32\times1$ k-mesh is employed for the
optimization of the lattice parameters and the ionic relaxation and a $48\times48\times1$
k-mesh for refining the electronic structure. We achieve an energy convergence of $10^{-7}$ eV
and a force convergence of 0.002 eV/\AA. Li-decorated monolayer graphene is modeled
by a $\sqrt3\times\sqrt3R30{^\circ}$ supercell with $a=b=2.26$ to that a Li atom
is added on each third hollow site. Phonon frequencies are determined by density functional
perturbation theory for evaluating the effects of the adatoms on the phonon spectrum, using
the scheme described in Ref.\ \cite{Mod}. The phonon dispersion is calculated with a
$24\times24\times1$ k-mesh. We study the effect of the substrate on the strength of the
electron-phonon coupling and the transition temperature for a supercell with Li-decorated 
monolayer graphene on top of $h$-BN with $a=b=4.32$ \AA\ and $c=15$ \AA\ (to avoid
artificial interaction due to the periodicity). Note that graphene on $h$-BN can be synthesized
due to the small lattice mismatch of only 1.4\% and interacts only weakly with the substrate
\cite{Dean,Xue,Yang}. By construction of the supercell of the suspended system, with 6 C atoms
and 1 Li atom, there are 21 phonon modes, whereas we have 39 modes for the supported system.

\subsection{Superconducting transition temperature}
The Allen-Dynes formula \cite{dynes,allen}, which is a modification of McMillan's
formula \cite{mcmillan}, is used to calculate
\begin{equation}
\label{equation1}
T_{c} = \frac{<\omega>_{\log}}{1.20} \exp\Big(-\frac{1.04 (1+\lambda)}{\lambda-\mu^*(1+0.62 \lambda)}\Big).
\end{equation}
The terms $<\omega>_{\log}$, $\lambda$, and $\mu^*$ are the logarithmic frequency average,
electron-phonon coupling constant, and effective Coulomb repulsion, respectively. Moreover,
the dimensionless parameter
\begin{equation}
\label{equation2}
\lambda = 2\int_0^\infty \frac{d\omega \alpha^2F(\omega)}{\omega}
\end{equation}
measures the strength of the Eliashberg function
\begin{equation}
\label{equation3}
\alpha^2F(\omega) = N_\uparrow(0)\frac{\sum_{kk'}|M_{kk'}|^2\delta(\omega-\omega_q)\delta(E_k)\delta(E_{k'})}{\sum_{kk'}\delta(E_k)\delta(E_{k'})},
\end{equation}
where $k$ and $q$ represent the electron band index and phonon wave number, respectively.
In addition, $N_\uparrow(0)$ is the single-spin density of states at the Fermi
surface and $M_{kk'}$ is the matrix element for electron-phonon coupling. The effective Coulomb
repulsion (also called Coulomb pseudopotential) is given by \cite{anderson}
\begin{equation}
\label{equation4}
\frac{1}{\mu^*} = \frac{1}{\mu} + \ln\left(\frac{\omega_{el}}{\omega_{ph}}\right),
\end{equation}
where $\omega_{el}$ is the plasma frequency and $\omega_{ph}$ the frequency cutoff in
$\alpha^2F(\omega)$. The Coulomb coupling $\mu$ is given by the product of the density
of states at the Fermi surface and the matrix element of the screened Coulomb interaction
averaged over the Fermi surface. We use $\mu^*=0.115$ in agreement with the experimental
observation of the critical temperature for bulk CaC$_6$ \cite{Emery}

\end{document}